\documentclass[superscriptaddress,twocolumn,showpacs,amsmath,amssymb]{revtex4}
\usepackage{graphicx}

\def\be{\begin{equation}}
\def\ee{\end{equation}}
\def\bea{\begin{eqnarray}}
\def\eea{\end{eqnarray}}

\renewcommand{\vec}[1]{\mbox{\boldmath $#1$}}

\begin{document}

\title{
Coulomb dissociation of the deformed halo nucleus $^{31}$Ne}

\author{Y. Urata}
\author{K. Hagino}
\affiliation{
Department of Physics,  Tohoku University,  Sendai,  980-8578,  Japan}

\author{H. Sagawa}
\affiliation{
Center for Mathematics and Physics,  University of Aizu,  Aizu-
Wakamatsu,  Fukushima 965-8560,  Japan}


\begin{abstract}
The recently observed large cross sections for the Coulomb dissociation 
of $^{31}$Ne nucleus indicate that this nucleus takes a halo structure in 
the ground state. 
We analyse these experimental data using the particle-rotor model, that 
takes into account the rotational excitation of the core nucleus $^{30}$Ne. 
We show that the experimental data can be well reproduced 
when the quadrupole deformation parameter of $^{30}$Ne is around 
$\beta_2=0.2\sim0.3$, at which the ground state takes the spin and parity 
of $I^\pi$=3/2$^-$. This state corresponds to the Nilsson level 
[330 1/2] in the adiabatic limit of the particle-rotor model. 
On the other hand, 
the state corresponding 
to the Nilsson level [321 3/2] with 
$\beta_2\sim 0.55$ can be excluded when the finite excitation energy 
of the core  is taken into account, even though this configuration is  a 
 candidate for the ground state of $^{31}$Ne 
in the Nilsson model analysis. 
We discuss briefly also a possibility of the $I^\pi$=1/2$^+$ configuration 
with $\beta_2\sim 1$ and $\beta_2\sim -0.4$. 
\end{abstract}

\pacs{21.10.Gv,21.60.-n, 25.60.Gc, 27.30.+t}

\maketitle

\bigskip

A halo structure with an extended density distribution is one of 
the characteristic features of weakly bound neutron-rich nuclei. 
It was first recognized in $^{11}$Li, showing an abnormally large 
interaction cross section \cite{T85}. 
Other examples include $^{11}$Be \cite{T88,F91} and $^{19}$C \cite{N99}, 
both of which are typical one-neutron halo nuclei. 
The root-mean-square radius diverges 
for $s$ and $p$ waves 
as the single-particle energy approaches to zero 
\cite{RJM92}, and the halo structure has been ascribed to an 
occupation of an $l=0$ or $l=1$ orbit by the valence neutron \cite{S92}. 

The halo structure induces a large concentration of the electric dipole (E1) 
strength in the low-excitation energy region, that is, a soft E1 excitation 
\cite{N99,N97,N06}. 
Recently, the Coulomb breakup cross sections for $^{31}$Ne were measured 
by Nakamura {\it et al}.\cite{N09}, indicating a soft E1 excitation 
in the $^{31}$Ne nucleus. 
Notice that a naive spherical shell model leads to the 1$f_{7/2}$ 
configuration for the valence neutron of $^{31}$Ne. In order to generate 
the halo structure within the mean-field picture, 
the valence neutron therefore needs to 
move in a deformed 
mean-field potential in which the $s$ or $p$ wave component in a 
weakly-bound single-particle wave function makes a dominant 
contribution \cite{MNA97,H04,YH05}. 
Based on this idea, Hamamoto has carried out Nilsson model calculations with 
a deformed Woods-Saxon potential, and argued that the large Coulomb 
dissociation cross sections of $^{31}$Ne can be understood if 
the valence neutron occupies [330 1/2], [321 3/2], or [200 1/2] Nilsson 
levels \cite{H10} leading to $I^{\pi}=3/2^-$ ([330 1/2] and 
[321 3/2] configurations) or $1/2^+$ ([200 1/2] configuration) for the spin 
and parity of the ground state of $^{31}$Ne. 
We mention that the ground state with $I^\pi=3/2^-$ is found also with 
shell model calculations \cite{N09,PR94} as well as with 
a microscopic cluster model calculation\cite{D99}. 

In order to describe odd-mass deformed nuclei, especially 
nuclei with small deformation as well as transitional nuclei, 
the particle-rotor model (PRM) has been applied with great success
\cite{RS80,BM75,BBH87}.  
It was also applied successfully to a halo nucleus $^{11}$Be  \cite{EBS95,NTJ96}
 and neutron-rich Carbon isotopes   $^{15,17,19}$C\cite{TH04}. 
In  $^{30}$Ne,  the candidates for the first excited $2^{+}$ and $4^{+}$ states 
have been identified experimentally  at excitation energies of 0.801 MeV and 
2.24 MeV, respectively \cite{D09,F10}. The energy ratio $E_{4^+}/E_{2^+}$=2.80 suggests 
this nucleus to be a transitional one in comparison with the ratio 3.33 for 
well-deformed nuclei.  Although the
experimental uncertainty is still large, 
an  extremely small neutron separation energy S$_n$=0.29$\pm$1.64 MeV\cite{J07} 
suggests  a halo structure of  $^{31}$Ne.   
Because of these reasons, in this paper we assume a deformed $^{30}$Ne core and 
adopt the particle-rotor 
 model to study the ground state and 
the excited state properties  of $^{31}$Ne. 

The Nilsson model adopted in Ref. \cite{H10} 
to analyze the ground state configuration of $^{31}$Ne 
corresponds to   
a strong coupling limit of the particle-rotor model. 
That is, the non-adiabatic effect 
due to the rotational states of the deformed core 
was neglected in the analysis.   
In general, in deformed nuclei with mass A$\lesssim 40$, the rotational energy 
is not small, and 
the non-adiabatic 
effect may play a role in discussing physical observables such as electric 
dipole (E1) transitions. 
While  the non-adiabatic effect may be estimated perturbatively in 
the Nilsson model, especially to determine the spin and parity of the 
 nucleus in the laboratory frame \cite{RS80,BM75}, it is not 
obvious whether the perturbation works for the particular case of 
$^{31}$Ne nucleus.  In this respect, it is highly desirable to 
disentangle how  much the non-adiabatic effect appears in the assignment of the spin-parity of 
the ground state  and also in the E1 transitions of $^{31}$Ne. 

The aim of this paper is thus to study the $^{31}$Ne nucleus by solving 
 the particle-rotor model Hamiltonian  
without resorting to the adiabatic approximation and discuss the E1 excitation 
and the Coulomb breakup cross sections. 
In order to describe the structure of the $^{31}$Ne nucleus, we assume that 
the core nucleus $^{30}$Ne is statically deformed with a quadrupole 
deformation parameter $\beta_2$ and the motion of the valence neutron is 
coupled to the rotational motion of the core nucleus. 
Assuming the axially symmetric shape of the core nucleus, 
we consider the following Hamiltonian for the $^{30}$Ne+$n$ system:
\begin{equation}
H = -\frac{\hbar^2}{2\mu}\vec{\nabla}^2+V(\vec{r},\hat{\vec{r}}_c)+H_{\rm rot},
\label{Hprot}
\end{equation}
where $\mu=m_{\rm N}A_c/(A_c+1)$ is the reduced mass of the valence neutron 
with $A_c$=30 and $m_{\rm N}$ being the mass number of the core nucleus 
and the nucleon mass, respectively. 
$H_{\rm rot}$ is the rotational Hamiltonian for the core nucleus 
given by $H_{\rm rot}=\vec{I}_c^2\hbar^2/2{\cal J}$, where 
$\vec{I}_c$ is the spin of the core nucleus and ${\cal J}$ is the moment of 
inertia. 
$V$ is the single-particle potential for the valence neutron interacting 
with the deformed core. $\vec{r}$ and $\hat{\vec{r}}_c$ 
are the coordinate of the 
valence neutron and the direction of the symmetry axis of the core nucleus 
in the laboratory frame, respectively. For the single-particle potential 
$V$, we use a deformed Woods-Saxon potential, 
\begin{eqnarray}
V(\vec{r},\hat{\vec{r}}_{c})&=&
-V_{\rm WS}\,\left(1-F_{\rm ls}r_0^2\,(\vec{l}\cdot\vec{s})\,
\frac{1}{r}\frac{d}{dr}\right)
f(r)\nonumber \\ 
&&+V_{\rm WS}R_0\beta_2\frac{df(r)}{dr}\,Y_{20}(\hat{\vec{r}}_{\rm cn}), \\
&\equiv& V_0(r)+V_2(r)Y_{20}(\hat{\vec{r}}_{\rm cn}), 
\end{eqnarray}
where $\hat{\vec{r}}_{\rm cn}$ is the angle between 
$\hat{\vec{r}}$ and $\hat{\vec{r}}_c$, and $f(r)$ is given by 
\begin{equation}
f(r)=1/\left[1+\exp((r-R_0)/a)\right],
\label{ws}
\end{equation}
with $R_0=r_0A_c^{1/3}$. 
Here, for simplicity, we have expanded the deformed Woods-Saxon potential 
up to the linear order of $\beta_2$. 
We have checked that this approximation works for the range of 
deformation parameter considered in this paper, by comparing our results 
for the Nilsson model with the numerical results obtained with the code 
{\tt WSBETA}\cite{CDNSW87}. 

In principle, the eigenfunctions of the Hamiltonian (\ref{Hprot}) 
can be obtained by solving the coupled-channels equations in the coordinate 
space both for the bound and the scattering states \cite{EBS95,TH04,ED00}. 
A continuum B(E1) distribution can then be constructed with those 
scattering wave functions. 
However, since the experimental data are so far available 
only in the form of inclusive cross sections, 
we  will not discuss the detailed structure of the 
strength distribution here but leave it for future publications.
It is sufficient for our purpose 
to expand the wave function on some basis and 
compute a discrete strength distribution. 
We do this with the eigenfunctions of the spherical part of 
the potential $V$, 
\begin{equation}
\left[-\frac{\hbar^2}{2\mu}\vec{\nabla}^2+V_0(r)-\epsilon_{njl}\right]
R_{njl}(r){\cal Y}_{jlm}(\hat{\vec{r}})=0,
\label{sphe}
\end{equation}
where $R_{njl}(r)$ is the radial wave function and 
${\cal Y}_{jlm}(\hat{\vec{r}})$ is the spin-angular function. 
The continuum spectrum can be discretized within a large box. 
Together with the rotational wave function $\phi_{I_cM_c}(\hat{\vec{r}}_c)$, 
the total wave function for the $n$+$^{30}$Ne system is expanded as, 
\begin{equation}
\Psi_{IM}(\vec{r},\hat{\vec{r}}_c)=\sum_{njl}\sum_{I_c}\alpha^{(I)}_{njlI_c}
R_{njl}(r)[{\cal Y}_{jl}(\hat{\vec{r}})\phi_{I_c}(\hat{\vec{r}}_c)]^{(IM)}, 
\label{wf}
\end{equation}
where $I$ is the spin of $^{31}$Ne and $M$ is its $z$-component. 
The expansion coefficients $\alpha^{(I)}_{njlI_c}$ as well as the 
corresponding eigenenergy for the $^{31}$Ne nucleus are obtained by 
numerically diagonalizing the Hamiltonian $H$. 

In order to identify the ground state configuration, we first solve the Hamiltonian 
by setting $H_{\rm rot}=0$ in Eq. (\ref{Hprot}). In this case, the $K$ quantum number, that is, 
the projection of the total 
angular momentum onto the $z$-axis in the body-fixed frame, is conserved, 
and several states with different $I$, having the same value of $K$, are degenerate in energy when 
the maximum value of $I_c$ included in the calculation is sufficiently large. 
As has been shown in Ref. \cite{ED00}, the wave function in this limit is related to the wave function 
in the Nilsson model, $\phi_{jlK}$, as, 
\begin{equation}
\sum_n\alpha^{(I)}_{njlI_c}R_{njl}(r)=A^{IK}_{jI_c}\phi_{jlK}(r),
\label{nilwf}
\end{equation}
where 
\begin{equation}
A^{IK}_{jI_c}=\sqrt{\frac{2I_c+1}{2I+1}}\cdot\sqrt{2}\,\langle j K I_c 0|I K\rangle, 
\end{equation}
and $\phi_{jlK}$ satisfies \cite{HG04} 
\begin{equation}
\left[-\frac{\hbar^2}{2\mu}\vec{\nabla}^2+V(\vec{r},\hat{\vec{r}}_c=0)-\epsilon_K\right]
\left(\sum_{j,l} \phi_{jlK}(r){\cal Y}_{jlK}(\hat{\vec{r}})\right)=0. 
\end{equation}
One can regard Eq. (\ref{nilwf}) as a transformation of the Nilsson 
wave function from the body-fixed frame to 
the laboratory frame, where the angular momentum is conserved. 
Notice that 
\begin{equation}
\sum_{I_c}\sum_n |\alpha^{(I)}_{njlI_c}|^2=\int r^2dr \phi_{jlK}(r)^2=P_{jl}^{\rm(Nil)}, 
\label{Pnil}
\end{equation}
is the probability of each $(j,l)$ component in the Nilsson wave function and is independent of $I$ 
when $H_{\rm rot}=0$. 

In this limit, we therefore obtain a collection of 
Nilsson levels. As usual, we put two neutrons to each Nilsson orbit from the 
bottom of the potential well, and seek the Nilsson orbit which is occupied by the last 
unpaired neutron. We then gradually 
increase the value of the 2$^+$ energy of the core nucleus up to the 
physical value, $E_{2^+}$=0.801 MeV, and 
monitor how the Nilsson orbit for the valence neutron evolves. 
For a finite value of $E_{2^+}$, the $K$ quantum number is not conserved 
any more due to the Coriolis coupling, and 
the degeneracy with respect to $I$ is resolved. 
We select the lowest energy state among several $I$ at 
$E_{2^+}$=0.801 MeV as the ground state of $^{31}$Ne. 
In this way, we take into account the Pauli principle between 
the valence neutron and the neutrons in the 
core nucleus. 

\begin{table*}[htb]
\caption{
The probability of the component [$I_c\otimes (jl)$] 
in deformed wave functions of $^{31}$Ne, where $I_c$ is the spin of the 
rotational state of the core nucleus $^{30}$Ne and $(jl)$ is the angular 
momentum between the valence neutron and the core nucleus. 
These are obtained with the particle-rotor model (PRM), where 
the results with vanishing rotational energy, $E_{2^+}$=0, 
correspond to the Nilsson model calculations. 
The states with $\beta_2$=0.1 and 0.2 correspond to the Nilsson orbit 
[330 1/2] while 
the state with $\beta_2$=0.55 corresponds to the Nilsson orbit 
[321 3/2]. The $K^\pi$ quantum number is therefore $K^\pi$=1/2$^-$ and 
$K^\pi$=3/2$^-$, respectively. The spin of $^{31}$Ne in the 
laboratory frame is $I^\pi=3/2^-$ for $\beta_2$=0.2 and 0.55 while 
it is $7/2^-$ for $\beta_2$=0.1. 
Two different values of the one neutron separation energy $S_n$
of the $^{31}$Ne nucleus are compared. 
For the case of $E_{2^+}=0$, we also list the total p$_{3/2}$ and 
f$_{7/2}$ probabilities 
defined as $P_{jl}\equiv \sum_{I_c}P_{jlI_c}$. 
}
\begin{center}
\begin{tabular}{l| c c c c c c c c}
\hline \hline
 & [0$^+\otimes$ p$_{3/2}$] & [2$^+\otimes$ p$_{3/2}$]
& (total p$_{3/2}$) & [2$^+\otimes$ p$_{1/2}$]  & [0$^+\otimes$ f$_{7/2}$]
 & [2$^+\otimes$ f$_{7/2}$] & [4$^+\otimes$ f$_{7/2}$] & (total f$_{7/2}$)\\ \hline
$S_n$=0.3 MeV & & & & & & &\\
$\beta_2$=0.1, ~~$E_{2^+}$ = 0 (Nilsson) & 0\% & 19.5\% & 30.3\% & 0\% & 17.2 \% & 20.5\% & 18.1 \% & 68.8\%\\
\hspace*{1.4cm}$E_{2^+}$ = 0.801 MeV (PRM) & 0\% & 6.5\% & - & 0 \% & 75.2\%
& 17.5 \% & 0.5 \% & - \\
$\beta_2$=0.2, ~~$E_{2^+}$ = 0 (Nilsson) & 22.2\% & 22.2\% & 44.4\% & 2.9\% & 0\% & 33.5 \% & 18.6 \% & 52.1\%\\
\hspace*{1.4cm}$E_{2^+}$ = 0.801 MeV (PRM) & 44.9\% & 8.4\% & - & 2.0\% & 0\% & 42.7 \% & 1.5 \% & - \\
$\beta_2$=0.55, $E_{2^+}$ = 0 (Nilsson) & 11.4\% & 11.5\% & 22.9\% & 0\% & 0\% & 25.5 \% & 45.9 \% & 71.4 \%\\
\hspace*{1.4cm}$E_{2^+}$ = 0.801 MeV (PRM) & 1.9\% & 29.7\% & - & 4.4\% & 0\% & 23.0 \% & 35.5 \% & -\\
\hline
$S_n$=0.1 MeV & & & & & & &\\
$\beta_2$=0.1, ~~$E_{2^+}$ = 0 (Nilsson) & 0\% & 26.7\% & 41.5\% & 0\% & 14.3\%
& 17.0\% & 15.0\% & 57.1\%\\
\hspace*{1.4cm}$E_{2^+}$ = 0.801 MeV (PRM) & 0\% & 7.2\% & - & 0\% & 74.8\%
& 17.1\% & 0.5\% & - \\
$\beta_2$=0.2, ~~$E_{2^+}$ = 0 (Nilsson) & 26.5\% & 26.5\% & 53.1\% & 3.8\% & 0\% & 27.4 \% & 15.2 \% & 42.6\%\\
\hspace*{1.4cm}$E_{2^+}$ = 0.801 MeV (PRM) & 53.4\% & 8.3\% & - & 2.1\% & 0\% & 34.6 \% & 1.2 \% & - \\
$\beta_2$=0.55, $E_{2^+}$ = 0 (Nilsson) & 12.9\% & 12.9\% & 25.7\% & 0\% & 0\% & 24.6 \% & 44.2 \% & 68.8 \%\\
\hspace*{1.4cm}$E_{2^+}$ = 0.801 MeV (PRM) & 1.8\% & 30.9\% & - & 4.9\% & 0\% & 23.6 \% & 33.5 \% & -\\
\hline \hline
\end{tabular}
\end{center}
\end{table*}

Table I shows the probability of each component, 
$P_{jlI_c}=\sum_n|\alpha^{(I)}_{njlI_c}|^2$, 
in the wave function for $\beta_2$=0.1, 0.2 and 
0.55 with two different values of the 
ground state energy of $^{31}$Ne. 
The ground state energy obtained by diagonalizing the Hamiltonian $H$ is 
measured from the threshold of $n$+$^{30}$Ne, and thus it is equivalent 
(except for the negative sign) 
to the one-neutron separation energy, $S_n$ 
(it should not be confused 
with the energy $\epsilon_{njl}$ in Eq. (\ref{sphe})). 
Following Ref. \cite{H10}, 
we use $r_0$=1.27 fm, $a$=0.67 fm, and $F_{\rm ls}$=0.44 for the parameters 
of the Woods-Saxon potential, (\ref{ws}). 
$V_{\rm WS}$ is adjusted to reproduce a given value of $S_n$. 
We use $R_{\rm box}$=60 fm for the size of the box to discretize the 
continuum spectrum, and include the single-particle orbits (\ref{sphe}) 
up to $\epsilon$ = 90 MeV and $l_{\rm max}$=10. 
For the spin of the rotational states of the core nucleus, we include 
up to $I_c=8$. 
In the case of vanishing rotational energy, 
the states with $\beta_2$=0.1 and 0.2 correspond to the Nilsson orbit 
[330 1/2] ($K^\pi=1/2^-$) while 
the state with $\beta_2$=0.55 corresponds to the Nilsson orbit 
[321 3/2] ($K^\pi=3/2^-$). See {\it e.g.,} Fig. 2 in Ref. \cite{H10} 
for the Nilsson diagram. 
The spin of $^{31}$Ne in the 
laboratory frame is $I^\pi=3/2^-$ for $\beta_2$=0.2 and 0.55, while it 
is $I^\pi=7/2^-$ for $\beta_2$=0.1. 

In the adiabatic limit (that is, the Nilsson model), 
for a given $j$ and $l$, 
the contribution of each component of [$I_c\otimes (jl)$] 
with different value of $I_c$ 
to the E1 excitation 
is similar to each other since the radial dependence of the 
wave function is identical (see Eq. (\ref{nilwf})). 
Therefore, the relevant quantity in this limit is the total 
probability for each $j$ and $l$ given by Eq. (\ref{Pnil}). 
On the other hand, in the case of finite rotational energy ({\it i.e.,} the non-adiabatic 
coupling), 
the radial dependence of the wave function largely depends on $I_c$. 
The wave function is spatially most extended for the component which couples to 
$I_c=0$, as the absolute value of the diagonal energy $E-E_{\rm rot}$ 
is the smallest (that is, most weakly bound). 
Therefore, the relevant quantity in this case to the E1 excitation is 
the probability of the $I_c=0$ component. 
In the case of $\beta_2=0.2$, the probability of [0$^+\otimes$p$_{3/2}$] 
increases considerably as the rotational energy increases from zero to 
0.801 MeV, and the probability at $E_{2^+}=0.801$ MeV is similar to the total p$_{3/2}$ probability 
in the adiabatic limit. As the separation energy decreases from 
$S_n$=0.3 MeV to $S_n$=0.1 MeV, 
the total $p$-wave probability increases in the adiabatic 
limit \cite{MNA97,H04,YH05,H10}, and correspondingly 
the probability of [0$^+\otimes$p$_{3/2}$] also increases in the 
non-adiabatic case. 
In contrast, in the case of $\beta_c=0.55$, the probability of 
[0$^+\otimes$p$_{3/2}$] decreases considerably as the rotational energy 
increases. The probability remains small even if the separation energy 
decreases from 
$S_n$=0.3 MeV to $S_n$=0.1 MeV. We have confirmed that this is the case 
even when the separation energy is as small as $S_n=0.01$ MeV. 
We thus expect that the dissociation cross section decreases significantly 
for this configuration due to the non-adiabatic effect. 
In the case of $\beta_2=0.1$, 
the [0$^+\otimes$f$_{7/2}$] component increases significantly 
for the non-adiabatic coupling. Since the $f$-wave does not form a halo structure, 
the E1 excitation probability will decrease considerably if the rotational energy 
is taken into account. As the separation energy decreases, the $p$-wave 
component will increase in the Nilsson wave function. However, 
the non-adiabatic effect always quenches the $p$-wave component in this 
weak coupling regime, and one could 
not expect large Coulomb dissociation cross sections with this configuration. 

Let us now numerically compute the Coulomb dissociation cross sections 
and confirm the above behaviors discussed in the previous paragraph. 
To this end, we first compute the E1 strength, 
\begin{equation}
B(E1; i\to f)=\frac{1}{2I_i+1}\left|\left\langle \Psi_f\left|\left|
\hat{D}\right|\right|\Psi_i
\right\rangle\right|^2,
\end{equation}
where the initial and final wave functions, $\Psi_i$ and $\Psi_f$, have the 
same form as in Eq. (\ref{wf}), and the dipole operator $\hat{D}_\mu$ is 
given by 
$\hat{D}_\mu=-[Z_c\,e/(A_c+1)]\cdot rY_{1\mu}(\hat{\vec{r}})$. 
The Coulomb dissociation cross sections can be obtained by 
multiplying the E1 virtual photon number, $N_{\rm E1}$, to the B(E1) strength 
\cite{EB92,BB88,WA79}. 
Summing all the final states, the inclusive Coulomb breakup cross section 
reads,
\begin{equation}
\sigma = \sum_f \frac{16\pi^3}{9\hbar c}\cdot N_{\rm E1}(E_f-E_i)\cdot 
B(E1; i\to f).
\label{cross}
\end{equation}
Since we compare our results with the experimental data for the dissociation 
cross sections, we restrict the summation in Eq. (\ref{cross}) only to 
those states above the threshold. 

\begin{figure}\label{fig1}
\includegraphics[scale=0.45,clip]{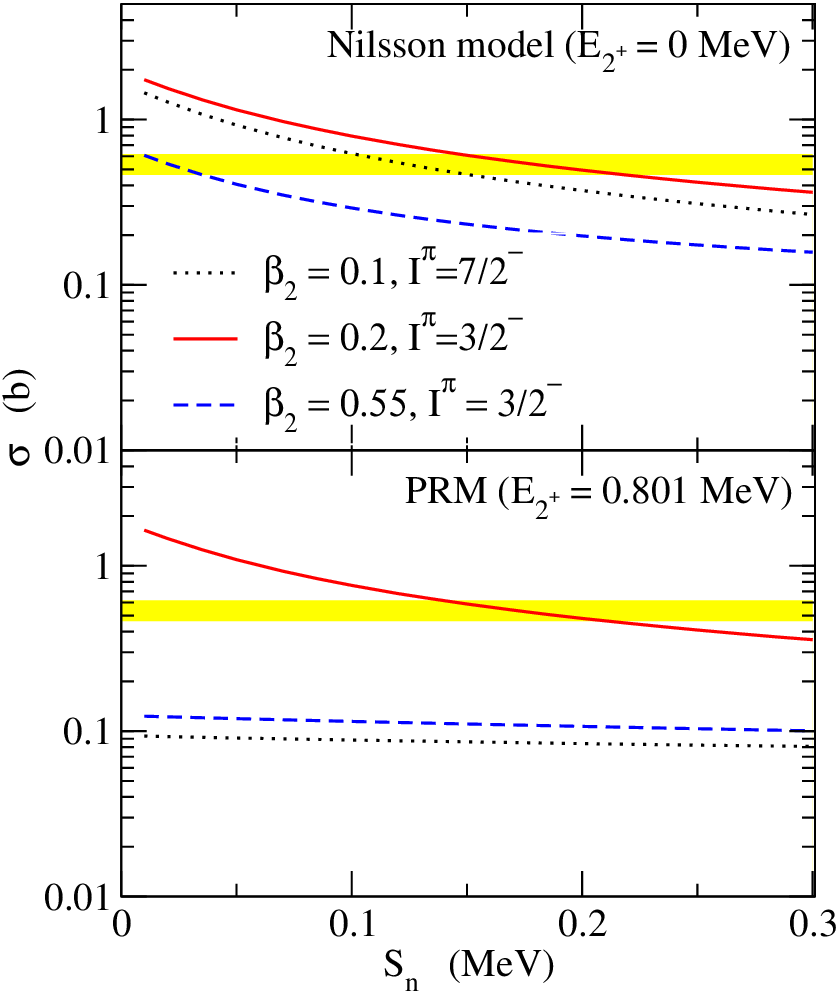}
\caption{(Color online)
The Coulomb dissociation cross sections of the $^{31}$Ne nucleus as a
function of the one-neutron separation energy $S_n$. 
The upper panel shows the results of the Nilsson model calculations, while 
the lower panel shows the results with a finite rotational energy of the 
core nucleus. The dotted, the solid, and the dashed lines denote 
the results with $\beta_2=0.1$, 0.2, and 0.55, respectively, where the 
spin and the parity of the ground state 
is $7/2^-$, $3/2^-$, and $3/2^-$, respectively. 
The shaded region indicates the experimental data \cite{N09}. 
}
\end{figure}

Figure 1 shows the Coulomb dissociation cross sections for the three 
configurations listed in Table I. Since the empirical separation energy 
$S_n$ has a large uncertainty, 
we plot the calculated cross sections as a function of 
$S_n$. The upper panel shows the results obtained in the adiabatic 
limit, while the lower panel shows the results with the finite rotational 
energy. The dotted, solid, and the dashed lines are obtained with 
$\beta_2$=0.1, 0.2, and 0.55, respectively. 
The shaded region indicates the experimental data \cite{N09}. 
One can clearly see
in the lower panel of Fig. 1 
that the experimental data can be well reproduced 
with $\beta_2=0.2$ if the separation energy $S_n$ is around 0.17 MeV. 
The results remain  almost the same even if we vary the deformation 
parameter in the range of $0.17 \lesssim \beta_2 \lesssim 0.33$. 
On the other hand, the cross sections obtained with $\beta_2$= 0.1 and 
0.55 appear too small to account for the experimental data. 
These configurations may yield a reasonable reproduction of the 
experimental data in the adiabatic limit because of the $p$-wave dominance 
in the wave function, but 
it is simply an artifact of the adiabatic approximation. 
The non-adiabatic effect eliminates the possibility of 
these configurations to be the ground state of $^{31}$Ne. 
All of these behaviors agree with the expectations. 

\begin{figure}
\includegraphics[scale=0.45,clip]{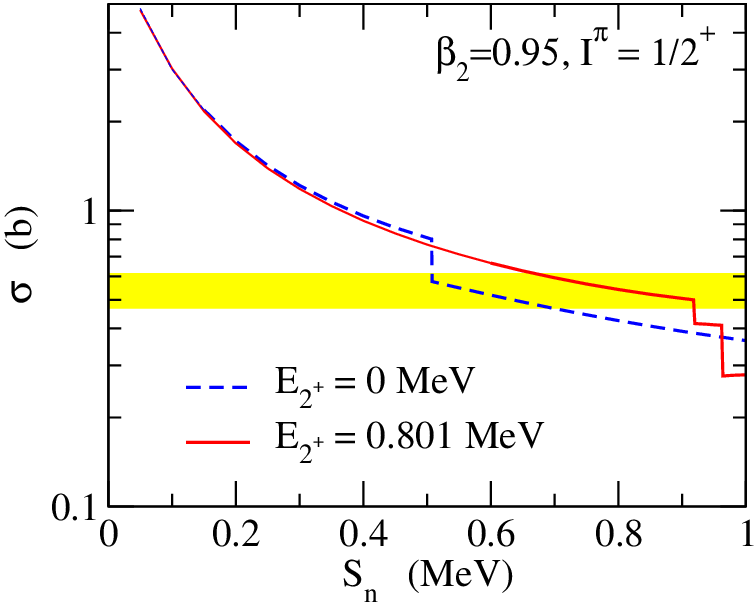}
\caption{(Color online)
The Coulomb dissociation cross sections of the $^{31}$Ne nucleus as a
function of the one-neutron separation energy $S_n$ at $\beta_2$=0.95. 
The corresponding Nilsson orbit is [200 1/2] ($K^\pi=1/2^+$), and the 
spin-parity of the ground state is $I^\pi=1/2^+$. 
The dashed and the solid lines denote 
the results in the adiabatic limit and in the non-adiabatic case, 
respectively. 
}
\end{figure}

Besides the configuration with $\beta_2\sim$0.2, there may be 
other possibilities for the ground state 
configuration of $^{31}$Ne. It was argued in Ref. \cite{H10} that 
the $I^\pi=1/2^+$ configuration originated from 
the [200 1/2] Nilsson orbit ($K^\pi=1/2^+$) cannot be excluded if the 
deformation parameter $\beta_2$ is large. 
Figure 2 shows the Coulomb dissociation cross sections 
calculated with $\beta_2=0.95$. The dashed and the solid lines correspond to 
the results in the adiabatic limit and those with the finite rotational energy, 
respectively. For this configuration, the cross sections show some jumps when 
a resonance state becomes a bound state as the separation energy increases 
so that its contribution is excluded 
in the sum in Eq. (\ref{cross}). 
In any case, the experimental cross sections can be reproduced when 
the separation energy is around 0.8 MeV, and thus this configuration cannot 
be excluded, although it is required that $^{31}$Ne has a 
bound excited state and/or low-lying resonance states. 
For this configuration, the [0$^+\otimes$s$_{1/2}$] configuration 
is the main component of the wave function both in the adiabatic limit and 
in the non-adiabatic case (for instance, the probability 
is 80.5 \% in the adiabatic limit and 79.2 \% in the non-adiabatic case), 
and the non-adiabatic effect on the Coulomb dissociation is found to be 
small. 

\begin{figure}
\includegraphics[scale=0.45,clip]{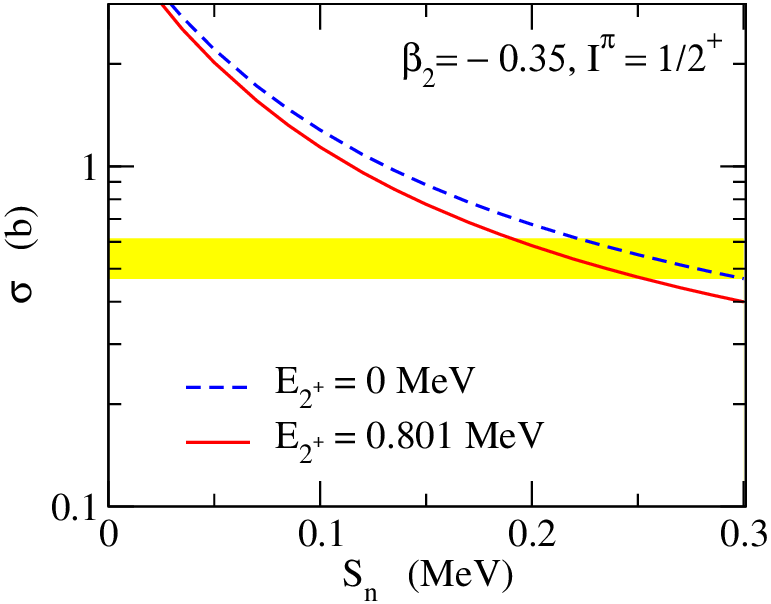}
\caption{(Color online)
Same as Fig. 2, but for $\beta_2=-0.35$. 
The corresponding Nilsson orbit is [200 1/2] ($K^\pi=1/2^+$), and the 
spin-parity of the ground state is $I^\pi=1/2^+$. 
}
\end{figure}

Although it may be unlikely that the $^{30}$Ne nucleus is oblately 
deformed \cite{F10}, the Coulomb dissociation data by themselves do not exclude a possibility of the 
oblate configuration corresponding to the [200 1/2] ($K^\pi=1/2^+$) 
Nilsson diagram 
at $\beta_2\sim -0.35$. In order to demonstrate this, Fig. 3 shows 
the calculated Coulomb dissociation cross sections obtained with $\beta_2=-0.35$. 
The spin-parity of the ground state is $I^\pi=1/2^+$. 
As one can see, the experimental Coulomb dissociation cross sections can be well 
reproduced when $S_n\sim$ 0.2 MeV. 
The main component of the wave function is the [2$^+\otimes$d$_{3/2}$] 
configuration with an appreciable mixture of the 
[0$^+\otimes$s$_{1/2}$] configuration. The $s$-wave probability does not 
change much by the non-adiabatic effect, and the experimental data can 
be reproduced both in the adiabatic limit and in the non-adiabatic case. 

In summary, 
we have discussed the E1 excitation of the $^{31}$Ne nucleus using the 
particle-rotor model. The finite rotational 
excitation energy of the core nucleus has been taken into account. 
We have shown that the experimental cross sections can be well 
reproduced with the deformation parameter of $0.17 \lesssim \beta_2 \lesssim 0.33$ 
and the separation energy of $0.13 \lesssim S_n \lesssim 0.2$ MeV, 
for which the ground state configuration has the spin-parity of 
$I^\pi=3/2^-$. However, 
the $I^\pi=1/2^+$ configurations with large deformation around 
$\beta_2\sim$ 0.95 or with oblate deformation around $\beta_2\sim -$ 0.35
cannot be excluded only from the Coulomb dissociation data. 
On the other hand,  
we have
 shown that the $I^\pi=7/2^-$ configuration at $\beta_2\sim 0.1$ 
and the $I^\pi=3/2^-$ configuration at $\beta_2\sim 0.55$ can be 
excluded due to the non-adiabatic effect, even though those configurations 
may reproduce the experimental data in the adiabatic limit. 

For the $^{31}$Ne nucleus, the momentum distribution for the nuclear breakup 
process \cite{HSCB10} as well as the interaction cross section have been 
recently measured at the RIBF facility at RIKEN \cite{N11,T10}. 
It will be interesting to investigate whether our model can reproduce 
these experimental data simultaneously. We will report on it in a separate 
publication. 

\bigskip

We thank T. Nakamura and K. Yoshida for useful discussions. 
This work was supported
by the Grant-in-Aid for Scientific Research (C), Contract No.
22540262 and 20540277 from the Japan Society for the Promotion of Science.

\end{document}